\documentclass[conference]{IEEEtran}
\IEEEoverridecommandlockouts
\usepackage{amsmath,amssymb,amsfonts}
\usepackage{algorithmic}
\usepackage{graphicx}
\usepackage{textcomp}
\usepackage{biblatex}
\usepackage{booktabs}
\usepackage{tabularx}
\usepackage{array}
\usepackage{lipsum}
\usepackage{comment}
\usepackage{hyperref}

\usepackage{tikz}
\usetikzlibrary{positioning,arrows.meta,fit}

\usepackage{xcolor}
\def\BibTeX{{\rm B\kern-.05em{\sc i\kern-.025em b}\kern-.08em
    T\kern-.1667em\lower.7ex\hbox{E}\kern-.125emX}}
\begin{document}

\title{A Survey Examining Neuromorphic Architecture in Space and Challenges from Radiation\\
\thanks{*Corresponding author}
}

\makeatletter
\newcommand{\linebreakand}{%
  \end{@IEEEauthorhalign}
  \hfill\mbox{}\par
  \mbox{}\hfill\begin{@IEEEauthorhalign}
}
\makeatother

\author{
    \IEEEauthorblockN{Jonathan Naoukin*}
    \IEEEauthorblockA{
    \textit{University of Texas at Austin}\\
    Austin, USA \\
    jnaoukin@utexas.edu}
    \and
    \IEEEauthorblockN{Murat Isik}
    \IEEEauthorblockA{
    \textit{Stanford University}\\
    Stanford, USA \\
    mrtisik@stanford.edu}
    \and
    \IEEEauthorblockN{Karn Tiwari}
    \IEEEauthorblockA{
    \textit{Indian Institute of Science, Bangalore}\\
    Bengaluru, India \\
    karntiwari@iisc.ac.in}

    \linebreakand 
}

\IEEEoverridecommandlockouts
\maketitle
\IEEEpubidadjcol
\begin{abstract}
Inspired by the human brain's structure and function, neuromorphic computing has emerged as a promising approach for developing energy-efficient and powerful computing systems. Neuromorphic computing offers significant processing speed and power consumption advantages in aerospace applications. These two factors are crucial for real-time data analysis and decision-making. However, the harsh space environment, particularly with the presence of radiation, poses significant challenges to the reliability and performance of these computing systems. This paper comprehensively surveys the integration of radiation-resistant neuromorphic computing systems in aerospace applications. We explore the challenges posed by space radiation, review existing solutions and developments, present case studies of neuromorphic computing systems used in space applications, discuss future directions, and discuss the potential benefits of this technology in future space missions.
\end{abstract}


\section{Introduction}

The advent of neuromorphic computing has ushered in a new era of computational possibilities, particularly in the demanding environment of outer space. This survey paper delves into the intricacies of neuromorphic computing architectures and their potential to enhance aerospace applications significantly. By emulating the human brain's neural structure and processing methods, neuromorphic systems offer a paradigm shift in computational efficiency and processing speed. Integrating Spiking Neural Networks (SNNs) and Field Programmable Gate Arrays (FPGAs) is central to this advancement, enabling high throughput rates and real-time operational capabilities critical for space exploration and satellite communication \cite{han2020hardware, wang2014fpga, guo2021toward, }.

The potential applications of neuromorphic computing in space are vast. They range from advanced cybersecurity measures safeguarding communication between spacecraft and ground control to sophisticated autonomous control systems that manage intricate tasks such as docking and maneuvering. The inherent efficiency of neuromorphic systems, characterized by their low power consumption and rapid processing abilities, makes them particularly suited for space missions' prolonged and isolated conditions \cite{ortiz2022towards, bersuker2018neuromorphic, isik2022design, izzo2022neuromorphic, huynh2022implementing}.

However, the space environment is abundant with challenges, the most formidable being radiation. Space radiation, including high-energy particles from the sun and cosmic rays from beyond our solar system, constantly threatens electronic systems. It can lead to myriad issues ranging from transient faults to permanent damage, undermining the reliability and functionality of space-borne computing systems \cite{fleetwood2000overview, wrbanek2020space, dilillo2022space}.

This paper provides a comprehensive overview of radiation effects on neuromorphic computing systems, drawing on a wealth of studies and real-world incidents that illustrate the vulnerability of classical electronic components to space radiation. Single Event Latch-ups (SELs) and Single Event Upsets (SEUs) are discussed, which result in catastrophic failures or subtle errors in computation. Neuromorphic systems must be protected from radiation's deleterious effects using radiation-hardened components and advanced shielding techniques, which have been developed.

Moreover, the paper addresses technical aspects of radiation shielding, such as the use of tantalum foil and design considerations for creating radiation-tolerant neuromorphic hardware \cite{yan2020memristors}. It also discusses the critical need for robust error correction protocols and fault-tolerant system design to enable neuromorphic systems to operate reliably in harsh environments. Neuromorphic computing holds great promise for space applications, offering a path toward more efficient, powerful, and resilient computing solutions. However, realizing this potential requires a concerted effort to overcome the significant challenge of radiation tolerance. This survey aims to provide a thorough understanding of the current state of neuromorphic computing in space, the challenges it faces, and the future directions for research and development in this exciting field \cite{kang2020application, bersuker2018neuromorphic, koons1999impact, taggart2017situ}.

\begin{figure}[h]
\centering
\includegraphics[width=0.4\textwidth]{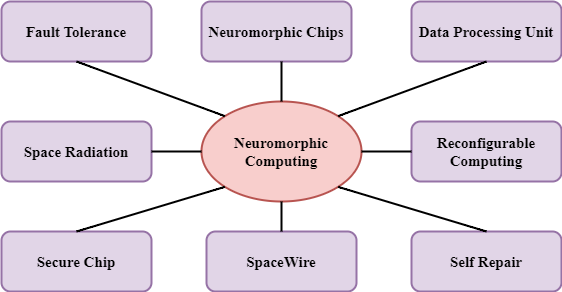}
\caption{Schematic layout of a neuromorphic computing system, highlighting its key components. The central unit, labeled "Neuromorphic Computing," symbolizes the core processing element, surrounded by various functional modules such as "Secure Chip," "Space Radiation," and "Fault Tolerance." The interconnection of those systems is depicted by dashed lines representing communication pathways essential for integrated operation. The peripheral components, "Data Processing Unit" and "Memory Storage," complete the system by providing data manipulation and retention capabilities. This abstract representation underscores the system's architecture designed for resilience and efficiency in processing neuromorphic algorithms.}
\label{fig:neuromorphic_computing_layout}
\end{figure}

~\autoref{tab:neuromorphic_space} below provides an overview of various organizations' use of neuromorphic computing technologies, highlighting the applications in space environments and the notable advantages and disadvantages.

\begin{table*}[ht]
\centering
\caption{Neuromorphic Computing in Space Radiation Environments}
\resizebox{\linewidth}{!}{%
\begin{tabular}{@{}llll@{}}
\toprule
\textbf{Advantage} & \textbf{Disadvantage} & \textbf{Application} & \textbf{Company/Organization} \\
\midrule
High Radiation Tolerance & Complexity in Design & Satellite On-board Processing & AFRL, Lockheed Martin \\
Low Power Consumption & Limited Commercial Adoption & Long-duration Space Missions & NASA, ESA \\
Fault Tolerant Design & Higher Initial Cost & Autonomous Spacecraft Operation & NASA, Lockheed Martin \\
Real-time Processing & Developmental Stage Technology & Real-time Data Analysis for Space Telescopes & ESA, NASA \\
Robustness to Single Event Upsets & Specialized Training Required & Robust Communication in Adverse Conditions & AFRL, Neurobus \\
Self-Repair Capabilities & Scalability Challenges & Adaptive Systems for Unpredictable Environments & Mentium Technologies, Neurobus \\
Energy Efficiency & Integration with Existing Systems & Energy-constrained Instruments and Rovers & NASA, ESA \\
High Computational Density & Limited Proven Use Cases & Compact Hardware for Spacecraft Instrumentation & Lockheed Martin, Mentium Technologies \\
\bottomrule
\end{tabular}%
}
\label{tab:neuromorphic_space}
\end{table*}

These organizations are at the forefront of integrating neuromorphic computing into their space programs to harness its potential for improving performance and reliability in radiation-rich environments. Neuromorphic computing presents a transformative potential for space applications by providing robust and energy-efficient solutions. However, the nascent nature of the technology poses challenges in design complexity, integration, and adoption. Leading companies ' ongoing research and development efforts aim to overcome these obstacles to harness the full potential of neuromorphic systems in demanding radiation environments.

\subsection{Contributions of Our Survey}

This manuscript comprehensively surveys the integration of radiation-resistant neuromorphic systems in aerospace applications. The major contributions of our survey are as follows:

\begin{itemize}
\item We provide an in-depth analysis of neuromorphic computing, detailing its inspiration from the human brain's structure, function, and potential advantages in space applications (particularly regarding processing speed and power consumption).
\item We explore the challenges posed by the harsh space environment, emphasizing the impact of radiation on neuromorphic computing reliability and performance.
\item We review existing solutions and developments in radiation-resistant neuromorphic computing, highlighting significant projects and research from organizations such as the European Space Agency (ESA) and National Aeronautics and Space Administration (NASA).
\item We present case studies or examples of neuromorphic systems that have been utilized in space applications, discussing how these systems have successfully addressed the challenge of radiation.
\item We analyze the future directions of radiation-resistant neuromorphic computing for aerospace applications, emphasizing the potential benefits and their impact on future space missions.
\end{itemize}

\subsection{Organization}
This manuscript is structured as follows: \textbf{Section II} provides a comprehensive background on neuromorphic computing and its relevance to aerospace applications. \textbf{Section III} delves into the challenges posed by space radiation on electronic systems, with a specific focus on neuromorphic systems. \textbf{Section IV} details the analysis of hardware reliability under the effect of radiation. \textbf{Section V} discusses existing solutions and developments in radiation-resistant neuromorphic computing, highlighting significant projects and research in the field. \textbf{Section VI} presents case studies or examples of neuromorphic computing systems that have been utilized in space applications, detailing how they have addressed the challenges of radiation. Finally, \textbf{Section VII} explores the future directions of radiation-resistant neuromorphic computing for space applications, emphasizing the potential benefits and impact on future space missions, and concludes the paper by summarizing critical points discussed throughout the manuscript.

\section{Background}

Neuromorphic computing is a field that synergizes the intricacies of biological systems with advancements in artificial intelligence. This architecture is becoming increasingly pivotal in aerospace engineering. This innovative computing paradigm, characterized by its emulation of the brain's neural architecture, offers staggering efficiency—up to 100,000 times more than traditional classical computers. This efficiency stems from integrating electronic hardware with numerous neurons and synaptic connections, akin to the human brain's functionality, a concept deeply rooted in the pioneering work of in the 1980s \cite{knowm_history_neuromorphic}.

Central to neuromorphic systems are artificial neurons and synapses, designed to process information similarly to biological neural networks. Inspired by the human brain, a complex system capable of performing diverse functions through interactions among billions of neurons and trillions of synapses, neuromorphic computing mimics the brain's adaptability in structure and function in response to environmental stimuli \cite{kenyon_mehonic_2022}.

SNNs, a crucial component of neuromorphic computing, processes information through artificial neurons that fire spikes or action potentials. These networks offer faster information processing capabilities than conventional digital networks because they do not require processing all inputs before generating outputs \cite{srag_nasa_gov, isik2023survey}.

The suitability of neuromorphic computing for aerospace applications is evident in its potential for spacecraft control, satellite communication, and rover navigation. The ability of these systems to learn and adapt to their environment by mirroring the human brain makes them particularly valuable for the dynamic and challenging conditions of space.

However, the space applications of neuromorphic computing face significant challenges because of space radiation. This radiation, which includes trapped radiation, solar particle events, and galactic cosmic rays, can cause severe damage to electronic systems. Radiation damages electronic components physically, causing SEUs, latch-ups, and other electrical malfunctions \cite{space_nss_org, nuclearsafety_gc_ca}.

Even with these challenges, neuromorphic computing is seeing practical applications in space, such as Neuro SatCOM. It has demonstrated the potential of neuromorphic computing to address complex problems in space \cite{connectivity_esa_int} using SNN Architecture in satellites.

Overall, neuromorphic computing is a technology well suited to space applications due to its power efficiency, enhanced computational power, and ability to simulate the human brain's adaptability. However, space radiation poses challenges that require ongoing research and development to ensure these systems are effective and reliable in this harsh environment \cite{ntrs_nasa_gov, degruyter_com}.

\section{Radiation Effects on Neuromorphic Computing}

Neuromorphic computing systems in space environments face significant radiation challenges. This causes data corruption, performance degradation, and even permanent damage to the system. Developing effective mitigation strategies requires an understanding of radiation's interaction with neuromorphic computing components. Throughout this section, we examine the physical interaction between radiation and neuromorphic components, typical radiation-induced failures, and the impact of different radiation types. Additionally, we discuss radiation thresholds and tolerances of neuromorphic components and how their architecture affects them.

\subsection{Physical Interaction of Radiation with Neuromorphic Components}

Radiation interacts with semiconductor materials in neuromorphic systems, potentially leading to errors. The Linear Energy Transfer (LET) equation is pivotal in understanding this interaction:
\begin{equation}
 LET = -\frac{dE}{dx}
\end{equation}
where \( dE \) represents the energy loss and \( dx \) represents the path length increment. This equation is crucial for understanding how radiation damages electronic components \cite{de2007radiation}.

\subsection{Radiation-Induced Failures in Neuromorphic Systems}

Neuromorphic computing systems are susceptible to radiation-induced failures, including SEUs, latch-ups, and total ionizing dose (TID) effects. The integral for TID effects measure the total ionizing dose absorbed by a material, expressed in Gray (Gy), which is one Joule/kg \cite{barnaby2006total}.

The equation for Total Ionizing Dose (TID) Effects is given by:
\begin{equation}
 TID = \int \Phi(E) \cdot S(E) \, dE
\end{equation}
Here, \( \Phi(E) \) is the flux of ionizing particles, and \( S(E) \) is the sensitivity of the material to ionizing radiation. This integral measures the total ionizing dose absorbed by a material.

\subsection{Cross-Section for Single-Event Effects (SEE)}
The cross-section for SEE is a function of the linear energy transfer and the material's properties used in neuromorphic components. It can be modeled as:
\begin{equation}
 Q = LET \times C \times s
\end{equation}
where \( Q \) is the charge collected in the sensitive volume, \( C \) is the capacitance of the sensitive volume, and \( s \) is the path length through the sensitive volume \cite{petersen2011single}.

\subsection{Bit Error Rate (BER) due to Radiation}
The Bit Error Rate (BER) due to radiation is an essential metric for estimating the rate of errors in a neuromorphic system due to radiation. It is expressed as the number of bit errors divided by the total number of bits \cite{detraz2009fpga}.

The equation for Bit Error Rate (BER) due to Radiation is given by:
\begin{equation}
 BER = \Phi \cdot \sigma_{SEE} \cdot N
\end{equation}
Where \( \Phi \) is the particle flux, \( \sigma_{SEE} \) is the cross-section for single-event effects, and \( N \) is the number of sensitive bits. This equation can estimate the rate of errors in a neuromorphic system due to radiation.

\subsection{Thresholds and Tolerances to Radiation}

Discussing the limits of radiation that neuromorphic components can withstand before malfunctioning or degrading is crucial. This involves understanding the thresholds beyond which components exhibit failure or significant performance degradation.

\begin{figure}[ht]
\centering
\includegraphics[width = 0.3\textwidth]{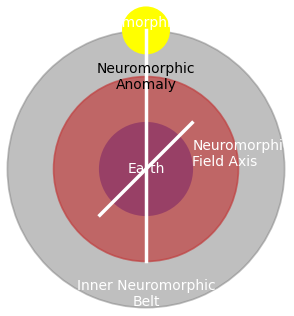}
\caption{Illustrative representation of the Earth's neuromorphic radiation belts and the Anomaly. The figure depicts the Earth at the center, surrounded by the inner and outer neuromorphic radiation belts in red and gray, respectively. The Anomaly is highlighted above the Earth, emphasizing its unique position. Additional annotations include the neuromorphic and the neuromorphic field axes, providing context for the orientation and directional properties relevant to neuromorphic computing systems in space.}
\label{Fig:1}
\end{figure}

Figure \ref{Fig:1} presents a comprehensive visualization of the Earth's neuromorphic radiation belts and the Anomaly, which is crucial for understanding the environmental challenges neuromorphic systems face in space. This graphical representation is meticulously designed to illustrate these phenomena' relative positions and scales with the Earth. At the center of the figure is a depiction of the Earth, rendered as a blue circle, symbolizing its physical location and size. It serves as a reference point for the surrounding features because of its central positioning. There are two distinct layers encircling the Earth, representing the inner and outer neuromorphic radiation belts. As a red semi-transparent wedge, the inner belt is indicated as being closer to the Earth and having a higher radiation intensity. Alternatively, the outer belt is depicted as a gray semi-transparent wedge, representing its extended reach beyond the inner belt. A yellow circle sits above the Earth, depicting the Anomaly in the figure. Anomaly's unique location and significance in studying radiation effects on neuromorphic systems are highlighted in this representation. An annotation is included for each element of the figure, which provides clarity and helps identify the Earth, the neuromorphic radiation belts, and the anomaly. By strategically placing these labels, we avoid obscuring the visual elements while also ensuring that they are readable. In addition, the figure delineates the neuromorphic axis and the neuromorphic field axis with white lines and text annotations. Orientation and directional properties relevant to this study are indicated by these axes. It is important to maintain an accurate aspect ratio when representing the relationship between the Earth's radiation belts, the Anomaly, and the radiation belts of the planet. The absence of traditional axes enables an unobstructed view of the depicted phenomena as the emphasis is placed on the graphic elements. The figure provides valuable insights into the environmental factors that affect neuromorphic systems in space by visualizing the spatial dynamics of the Earth's neuromorphic radiation belts and the Anomaly.

\subsection{Architecture's Influence on Radiation Tolerance}

The design and architecture of neuromorphic systems, such as redundancy and error correction, play a crucial role in mitigating radiation damage. This section explores how architectural choices influence the overall resilience of neuromorphic systems against radiation \cite{petersen2011single}.

Neuromorphic computing systems that can withstand radiation-rich environments and harsh conditions in space require a comprehensive understanding of these factors. Radiation challenges can be analyzed and addressed with the equations and models discussed. Radiation has detrimental effects on neuromorphic systems, causing data corruption, performance degradation, and even permanent damage. Understanding the interaction of radiation with neuromorphic components is crucial for developing effective mitigation strategies. These interactions lead to Total Ionizing Dose (TID) Effects, Single-Event Effects (SEEs), and Displacement Damage Effects \cite{easley1962radiation, braunig1999radiation}.

\subsection{Mechanisms of Radiation Interaction}
Radiation interacts with neuromorphic computing components through various mechanisms:

\begin{itemize}
    \item \textbf{Direct Ionization:} High-energy photons or charged particles such as electrons, protons, and heavy ions interact with a semiconductor material, generating electron-hole pairs along the particle's path due to energy loss, known as Linear Energy Transfer (LET) \cite{easley1962radiation}.
    
    \item \textbf{Indirect Ionization:} This mechanism focuses on damage to the crystal lattice of the semiconductor material, primarily caused by neutrons \cite{braunig1999radiation}.
    
    \item \textbf{Nuclear Reactions:} Incident protons can cause nuclear reactions in the semiconductor material, emitting secondary particles that cause additional ionization and displacement damage \cite{bird2017neutron}.
\end{itemize}

\subsection{Radiation-Induced Failures in Neuromorphic Systems}
Neuromorphic computing systems are susceptible to various radiation-induced failures:

\begin{itemize}
    \item \textbf{Single-Event Upsets (SEUs):} SEUs are changes of state caused by ionizing particles striking a sensitive node in a microelectronic device. In neuromorphic computing, SEUs can cause bit flips in memory elements \cite{bird2017neutron}.
    
    \item \textbf{Latch-ups:} A latch-up refers to a short circuit in an integrated circuit, disrupting the functioning of the neuromorphic system \cite{park2021integrate}.
    
    \item \textbf{Total Ionizing Dose (TID) Effects:} TID effects degrade the gain and speed of semiconductor devices, affecting the reliability and functionality of microelectronics \cite{brewer2019impact}.
    
    \item \textbf{Displacement Damage:} This refers to the displacement of atoms from the crystal lattice by incident radiation, affecting the device's electrical properties \cite{alenaradiation}.
\end{itemize}

\subsection{Effects of Different Radiation Sources}
Different radiation sources, such as solar flares, cosmic rays, and the Van Allen belts, have varying impacts on neuromorphic systems:

\begin{itemize}
    \item \textbf{Solar Radiation:} Solar wind, consisting of energetic electrons and protons, can cause TID effects and SEUs in neuromorphic systems \cite{hassan2023novel}.
    
    \item \textbf{Cosmic Radiation:} Galactic Cosmic Rays (GCRs), including alpha particles, protons, and high atomic number ions, are highly ionizing and particularly damaging \cite{molin2023neuromorphic}.
    
    \item \textbf{Van Allen Belts:} In Low Earth Orbit (LEO), the radiation dose can vary significantly, affecting the radiation tolerance requirements for neuromorphic computing components \cite{amarnath2023error}.
\end{itemize}

\subsection{Thresholds and Tolerances to Radiation}
The thresholds and tolerances of neuromorphic components to radiation vary depending on the component type and design. Radiation-hard devices can tolerate 100K to 1 MRad of TID, while radiation-tolerant devices can tolerate at least 30 KRad of TID. Commercial devices generally tolerate between 15 – 50 KRad of TID \cite{parlevliet2021autonomous}.

\subsection{Influence of Architecture on Radiation Tolerance}
The architecture of neuromorphic systems significantly influences their radiation tolerance:

\begin{itemize}
    \item \textbf{Device Technology and System Design:} RRAM-based neuromorphic systems are highly resistant to transient single-event effects and tolerant to multi-Mrad levels of total ionizing and displacement damage dose \cite{brewer2019impact}.
    
    \item \textbf{Error Correction Techniques:} These techniques detect and correct errors caused by radiation, enhancing system reliability \cite{hassan2023novel}.
    
    \item \textbf{Redundancy:} Incorporating redundancy at various levels ensures continued operation if one component fails due to radiation damage \cite{amarnath2023error}.
    
    \item \textbf{Adaptive and Self-healing Mechanisms:} These mechanisms adjust system parameters or reconfigure the architecture in response to radiation-induced changes, enhancing radiation tolerance \cite{alenaradiation}.
\end{itemize}

\section{Analysis of Hardware Reliability Under Radiation Exposure}

The reliability of hardware components in the presence of radiation is a paramount concern in designing and operating systems destined for environments with high levels of ionizing radiation, such as outer space. The Mean Time Between Failures (MTBF) metric is a pivotal indicator of the robustness and longevity of these components under such stressors \cite{tambara2016analyzing}.

\autoref{fig:mtbf_radiation} depicts a comparative study of various hardware systems, including Central Processing Units (CPUs), Graphics Processing Units (GPUs), FPGAs, and Neuromorphic Hardware. The analysis presents the normalized MTBF values under simulated radiation conditions, using Neuromorphic Hardware as the reference for normalization due to its cutting-edge resistance to radiation-induced failures.

\begin{figure}[ht]
\centering
\includegraphics[width=0.4\textwidth]{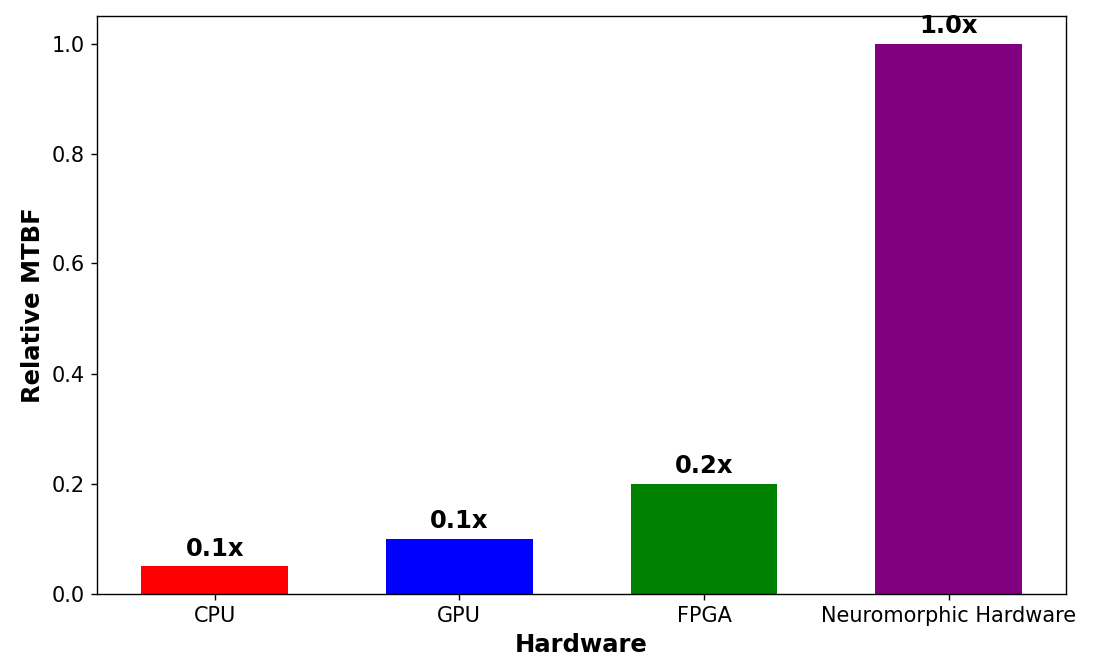}
\caption{Bar chart illustrating the relative MTBF under radiation for various hardware systems.}
\label{fig:mtbf_radiation}
\end{figure}

The methodology employed in this study includes:

\begin{itemize}
\item \textbf{Radiation Testing:} Exposing each hardware category to controlled radiation environments to emulate potential real-world conditions. These environments are generated using gamma rays and neutrons to ascertain the hardware's response to ionizing and non-ionizing energy transfers.
\item \textbf{Accelerated Life Testing:} Implementing stress testing at elevated radiation levels to precipitate the occurrence of failures, thus allowing for a timely assessment of the hardware's durability.
\item \textbf{Failure Monitoring:} Continuously tracking the performance of the hardware to identify the onset and frequency of failure, which contributes to the accurate calculation of MTBF.
\item \textbf{Statistical Analysis and Normalization:} Applying statistical models to predict MTBF and normalizing the data to ensure comparability across different hardware platforms.
\item \textbf{Modelling and Simulation:} Utilizing computational models to simulate radiation's long-term effects and forecast the gradual degradation of the hardware's integrity.
\end{itemize}

As demonstrated in the figure, the results indicate that Neuromorphic Hardware exhibits superior resilience with a significantly higher MTBF, denoting a lower frequency of failure events per operational hour under equivalent radiation levels. This underscores the potential of Neuromorphic Hardware in applications where reliability is critical under high-radiation conditions.

\section{Solutions and Developments}

Several solutions and developments have been made in radiation-resistant neuromorphic computing. Projects and research conducted by organizations such as ESA, NASA, and Connectivity ESA have focused on developing radiation-hardened electronic components and systems for space applications. These developments aim to improve the reliability and performance of neuromorphic systems in the harsh space environment. However, there are still gaps and limitations in the current solutions, such as the need for more robust and reliable radiation-hardened components. Integrating neuromorphic computing into aerospace applications presents a unique set of challenges, chief among them being the need to develop systems that can withstand the harsh radiation environment of space. The ESA, NASA, and the Connectivity ESA project have spearheaded research into creating neuromorphic systems that are not only efficient and powerful but also radiation-resistant. For example, recent ESA projects have demonstrated advancements in integrating silicon carbide-based components, which offer enhanced radiation resistance compared to traditional silicon-based systems \cite{gonzalez2012characterizing, goiffon2019radiation, barnaby2006total}.

\subsection{Radiation-Hardened Neuromorphic Chips}

The development of radiation-hardened neuromorphic chips represents a critical advancement in space-grade computing. These chips are meticulously engineered to withstand the severe ionizing radiation prevalent in the space environment, which can induce catastrophic failures in standard semiconductor devices. The hardening process is multifaceted, encompassing both innovative design strategies and the adoption of manufacturing techniques that leverage radiation-resistant materials \cite{dodd2003basic, winokur1994radiation, johnston2000radiation}.

\subsubsection{Design Strategies for Radiation Hardening}
The design of radiation-hardened neuromorphic chips often involves insulating substrates that prevent the formation of parasitic transistors, which can be triggered by radiation. Additionally, guard rings are implemented to isolate sensitive areas of the chip, providing an additional layer of protection against radiation-induced charge collection. These design strategies are complemented by redundant circuitry, which allows the chip to maintain operational integrity even if some parts are damaged. For instance, recent developments have seen the implementation of advanced error correction algorithms that enhance the resilience of these chips against transient radiation effects.

\subsubsection{Material Innovations}
The choice of materials in the fabrication of neuromorphic chips is pivotal for radiation hardening. Radiation-resistant chips have been made possible by using silicon-on-insulator (SOI) technology. Among the wide-bandgap semiconductors that offer superior radiation hardness to silicon are silicon carbide (SiC) and gallium nitride (GaN). Recently, these materials have shown promising results in tests, demonstrating enhanced durability in high-radiation environments.

\subsubsection{Testing and Validation}
Radiation-hardened neuromorphic chips must undergo rigorous testing under simulated space conditions to prove their effectiveness. A controlled laboratory environment is used to expose the chips to gamma rays, protons with high energies, and heavy ions. Further design and material refinements are made based on the data collected from these tests, ensuring that the chips can withstand space-related conditions. NASA's recent tests revealed critical insights into the performance of SOI-based chips under extreme radiation, leading to significant design improvements.

\subsubsection{Challenges and Trade-offs}
While significant progress has been made, there are inherent challenges and trade-offs in developing radiation-hardened neuromorphic chips. One of the primary challenges is maintaining the balance between radiation hardness and other performance metrics, such as power efficiency and processing speed. Additionally, the increased production costs associated with specialized materials and design complexities must be carefully managed to ensure the viability of these chips for widespread use in space missions. Balancing these factors remains a key focus area in ongoing research and development efforts.

\subsection{Radiation-Tolerant Architectures}

The pursuit of radiation-tolerant neuromorphic architectures extends beyond the resilience of individual chips to encompass the robustness of the entire computing system. This holistic approach to design ensures that neuromorphic systems can maintain operational integrity even in the face of intense radiation events that are common in extraterrestrial environments \cite{weigand2002radiation, sterpone2005analysis, kastensmidt2006fault}.

\subsubsection{System-Level Redundancy}
At the heart of radiation-tolerant architectures lies the principle of redundancy. This is realized through various strategies, such as duplicating critical pathways, which allows a system to continue functioning even when one part is compromised. Triple Modular Redundancy (TMR) is a common technique where three identical circuits perform the same operation, and a majority voting system decides the correct output. This approach is particularly effective in mitigating the impact of SEUs, which can alter the state of a bit within a circuit. Recent advancements in TMR design have led to more efficient and compact implementations, making them more feasible for space applications where size and weight are critical constraints.

\subsubsection{Error Detection and Dynamic Reconfiguration}
Radiation-tolerant architectures are characterized by error detection and dynamic reconfiguration. Radar-induced errors are continuously monitored by sensors and algorithms in these systems. A dynamically reconfigurable architecture can bypass the affected areas once it detects them, preserving its functionality. This adaptability is crucial for long-duration space missions with limited maintenance and repair options. For instance, a recent innovation in this area involves using machine learning algorithms to predict and preemptively address potential failure points, enhancing the system's resilience against unpredictable radiation events \cite{}.

\subsubsection{Radiation-Aware Algorithms}
The development of radiation-aware algorithms represents a novel approach to enhancing the radiation tolerance of neuromorphic systems. These algorithms are designed to be inherently robust against radiation-induced disruptions, often by incorporating fault-tolerant coding and data processing techniques. Integrating such algorithms into neuromorphic architectures ensures the system can maintain accurate and reliable operation even in radiation-induced noise and errors. Ongoing research in this domain explores the use of advanced neural network models that can adapt to and compensate for the effects of radiation, thereby maintaining the integrity of the computational processes.

\subsubsection{Challenges in Architecture Design}
Designing radiation-tolerant architectures involves navigating a complex landscape of challenges, particularly in balancing the competing demands of Size, Weight, and Power (SWaP) constraints with the need for robust radiation protection. Integrating additional protective features often increases system size and power consumption, which can be prohibitive for space applications where resources are limited. A recent case study involving the design of a satellite-based neuromorphic system highlighted the difficulties in achieving an optimal balance between SWaP requirements and radiation tolerance. The study underscored the need for innovative design approaches to reconcile these competing factors effectively.

\subsection{Self-Repairing Systems}

The concept of self-repairing systems in neuromorphic computing represents a paradigm shift towards creating autonomous, resilient computing architectures capable of withstanding the harsh conditions of space. A real-time fault detection and correction system ensures uninterrupted operation without external intervention \cite{ibe2010impact, reynolds2009quantitative, ban2012methods}.

\subsubsection{Autonomous Fault Detection and Real-Time Response}
Autonomous fault detection is at the core of self-repairing systems. The system's health and performance are continuously monitored using sensors and diagnostic algorithms. Real-time corrective actions can be initiated by the system when an anomaly is detected, such as a radiation-induced fault. Rerouting data pathways, activating redundant components, or adjusting operational parameters may be necessary to mitigate the impact of the fault. Recently, advanced diagnostic algorithms have been developed that can differentiate between transient and permanent faults, enabling more targeted and effective responses.

\subsubsection{Machine Learning for Predictive Maintenance}
A new avenue for predictive maintenance has been opened by integrating machine learning techniques into self-repairing systems. Analysis of system performance data patterns can help machine learning algorithms identify potential failure points before they occur. A proactive maintenance approach reduces the likelihood of system downtime and extends the lifespan of neuromorphic systems. Recently, a satellite-based neuromorphic system used machine learning algorithms to predict and prevent several potential failures, demonstrating the effectiveness of this approach.

\subsubsection{Challenges and Considerations}
Although self-repairing systems offer significant advantages in terms of resilience and autonomy, they also present unique challenges. The balance between the complexity of the self-repair mechanisms and the overall efficiency of the system is a primary consideration. In addition to increasing the power consumption and computational overhead of a system, adding fault detection and correction layers can negatively impact its performance. Moreover, erroneous responses to false positives can lead to unnecessary system disturbances if the self-repair algorithms are not reliable and accurate. The deployment of a self-repairing neuromorphic system on a Mars rover highlighted these challenges, emphasizing the need for rigorous testing and validation to ensure the system's reliability.

\subsection{Collaborative Efforts and Challenges}

Radiation-resistant neuromorphic systems require collaboration between various fields, including materials science, computer engineering, and space science. It is essential to use a collaborative approach when developing systems that can withstand extreme conditions in space \cite{jorgensen1991radiation, johnston2001space, wrbanek2020space}.

\subsubsection{Interdisciplinary Research and Collaboration}
Various domain experts contribute their skills and expertise to this interdisciplinary research because of its interdisciplinary nature. Materials scientists develop new radiation-hardened materials, while computer engineers integrate these materials into efficient and robust neuromorphic architectures. Collaboration with space agencies provides valuable insights into space applications' specific requirements and challenges, ensuring the developed systems are well-suited for their intended environment. It is evident that collective effort can overcome complex technical challenges, as evidenced by recent collaborations between academia and industry.

\subsubsection{Testing Rigors and Balancing Competing Demands}
Radiation-resistant neuromorphic systems require rigorous testing to ensure their reliability and performance in space, which is a critical aspect of their development. During this process, the systems are subjected to tests that simulate the harsh conditions of space, including extreme temperatures, vacuums, and high radiation levels. Balancing the competing demands of performance, size, weight, power consumption, and radiation resistance is a continual challenge. For example, a recent joint project between NASA and a leading university focused on optimizing the SWaP characteristics of a neuromorphic system for a lunar exploration mission. The project highlighted the intricate trade-offs in achieving an optimal balance between these competing factors.

\subsubsection{Future Outlook}
The radiation-resistant neuromorphic computing field holds immense potential for advancing space exploration and other high-radiation applications. The ongoing collaborative efforts and research pave the way for more resilient, efficient, and autonomous computing systems that can thrive in the most challenging environments. As technology continues to evolve, the boundaries of what is possible in this domain continually expand, offering exciting prospects for the future of space-grade computing.

\section{Case Studies or Examples}

The practical application of neuromorphic computing in space exploration has yielded several instructive case studies. These real-world examples act as beacons, guiding the development of future systems and providing empirical evidence of the efficacy of various radiation-resistant strategies.

\begin{figure}[h]
\centering
\includegraphics[width=0.4\textwidth]{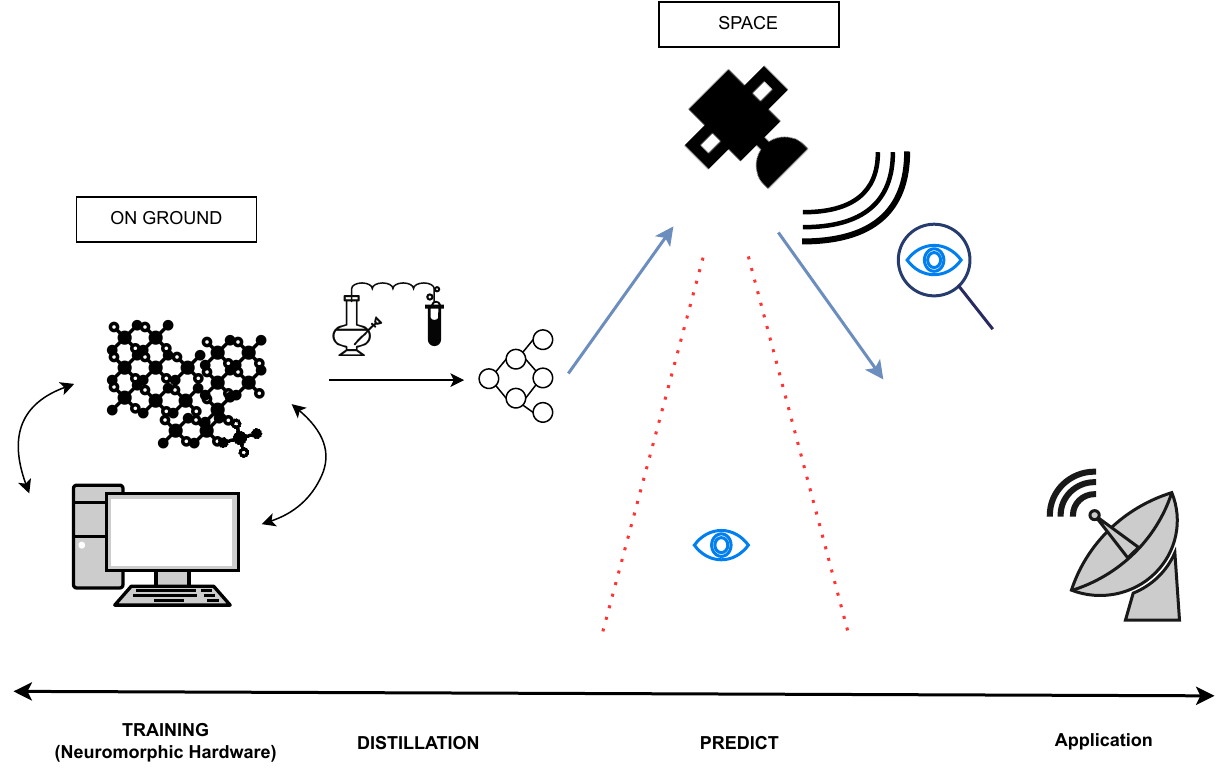}
\caption{Schematic workflow of neuromorphic hardware utilization in space applications. The process begins with 'Distillation,' where essential computational paradigms are extracted. This is followed by 'Training,' typically conducted on Earth, to adapt the neuromorphic systems for predictive tasks in space ('Predict'). This demonstrates the application of neuromorphic hardware in space, where satellite systems can leverage their inherent efficiency and fault tolerance, particularly when exposed to cosmic radiation and other extraterrestrial variables.}
\label{Fig:Neuromorphic_Space_Application}
\end{figure}

\subsubsection{Mars Rovers and Neuromorphic Vision}
The Mars rovers, equipped with neuromorphic vision systems, exemplify the successful application of these technologies in a space environment. These systems use radiation-hardened components and sophisticated error-correction algorithms to process visual data despite the constant radiation threat efficiently.

\subsubsection{Satellite Communication Systems}
Neuromorphic computing has also been integrated into satellite communication systems, where rapid and accurate signal processing is crucial. These systems demonstrate the use of fault-tolerant architectures that maintain communication links even when individual components are compromised by radiation.

\subsubsection{International Space Station Experiments}
Experiments on the International Space Station (ISS) provide valuable insights into the behavior of neuromorphic systems in low Earth orbit, an environment fraught with radiation challenges. These studies aid in refining the designs of neuromorphic systems, ensuring their viability for longer and more distant space missions.

\subsubsection{Deep Space Probes}
Deep space probes like those sent to the outer planets rely on neuromorphic computing for autonomous decision-making and data processing. These case studies emphasize the systems' ability to operate independently, far from the immediate support of ground control, and under intense cosmic radiation.

\subsubsection{Challenges Highlighted by Case Studies}
While these case studies underscore the progress, they also highlight ongoing challenges. For instance, the limited ability to repair or replace systems once deployed in space necessitates near-perfect reliability. Additionally, these case studies often reveal unforeseen interactions between radiation and other environmental factors, prompting further research and innovation.

\section{Future Directions and Conclusion}

As we venture further into the cosmos, the integration of radiation-resistant neuromorphic computing in aerospace applications is poised to play a pivotal role in the evolution of space missions. The potential of these systems to revolutionize data processing capabilities, inherent reliability, and energy-efficient nature positions them as an indispensable asset for prolonged extraterrestrial explorations. Future research should focus on refining the resilience of neuromorphic systems against the multifaceted challenges posed by space radiation. This includes developing innovative shielding techniques, error correction protocols, and adaptive algorithms that can mitigate the adverse effects of radiation, thereby ensuring the integrity and functionality of the computing systems in the harsh space environment.

Based on the surveyed related works and the current state of the field, we propose the following potential topics for further research in the area of radiation-resistant neuromorphic computing for aerospace applications:

\begin{itemize}
\item \emph{\textbf{Advanced Shielding Materials and Techniques:}} Investigate and develop innovative shielding materials and techniques that can provide robust protection against various forms of space radiation, ensuring the integrity and functionality of neuromorphic systems in the harsh space environment \cite{nasa_shielding}.

\item \emph{\textbf{Error Correction Protocols and Adaptive Algorithms:}} Explore the potential of error correction protocols and adaptive algorithms in mitigating the effects of radiation on neuromorphic systems, thereby enhancing their reliability and performance in space applications \cite{goiffon2019radiation}.

\item \emph{\textbf{Novel Materials and Architectures:}} Conduct extensive research on novel materials and architectures that are inherently resistant to radiation damage, paving the way for the development of more resilient neuromorphic systems \cite{barnaby2006total}.

\item \emph{\textbf{Artificial Intelligence and Machine Learning Integration:}} Examine the integration of artificial intelligence and machine learning algorithms to enhance the self-healing and adaptive capabilities of neuromorphic systems, thereby improving their resilience against radiation-induced damage \cite{james2017historical, huynh2022implementing}.

\item \emph{\textbf{Aerospace Applications of Neuromorphic Computing Systems:}} Investigate the potential of neuromorphic systems in specific aerospace applications, such as autonomous navigation, real-time data analysis, and decision-making, to fully harness their advantages in space missions \cite{furber2016large}.

\item \emph{\textbf{Cybersecurity Measures for Spacecraft Communication Systems:}} Explore the potential of neuromorphic systems in enhancing cybersecurity measures for communication systems between spacecraft and ground control, ensuring secure and reliable communication during space missions\cite{moskowitz2003self}.

\item \emph{\textbf{Autonomous Control Systems for Spacecraft Docking:}} Investigate the potential of neuromorphic systems in improving autonomous control systems for spacecraft docking and supply loading, thereby enhancing the efficiency and safety of space missions \cite{damico2005proximity}.

\end{itemize}

Furthermore, exploring novel materials and architectures inherently resistant to radiation damage should be a priority. Integrating artificial intelligence and machine learning algorithms to enhance these systems' self-healing and adaptive capabilities will also be a significant area of interest. Radiation-resistant technologies and neuromorphic computing represent a frontier in aerospace. The potential benefits of these systems, including enhanced data processing capabilities, improved reliability, and reduced power consumption, are integral to the success of future space missions. As we continue to push the boundaries of human knowledge and technology, the development and integration of radiation-resistant neuromorphic systems will undoubtedly be a cornerstone in our quest to explore the vast expanse of space.

\printbibliography

\end{document}